\documentclass[11pt,twoside]{article}
\usepackage{asp2004}
\usepackage{psfig}
\usepackage{epsf}
\usepackage{graphics}
\usepackage{lscape}
\def\edcomment#1{\iffalse\marginpar{\raggedright\sl#1\/}\else\relax\fi}
\reversemarginpar

\begin{document}
\title{Local Structures as Revealed by HIPASS}
\author{B. S. Koribalski}
\affil{Australia Telescope National Facility, CSIRO, P.O. Box 76, 
  Epping, NSW 1710, Australia}

\begin{abstract}
The HI Parkes All-Sky Survey (HIPASS) gives an unprecedented view of the
local large-scale structures in the southern sky. I will review the results
from the HIPASS Bright Galaxy Catalog (BGC, Koribalski et al. 2004) and the 
first version of the deep catalog (HICAT, Meyer et al. 2004) with particular 
emphasis on galaxy structures across the Zone of Avoidance. Some previously 
hardly noticed galaxy groups stand out quite distinctively in the HI sky 
distribution and several large-scale structures are seen for the first time.
Because HI surveys preferentially detect late-type spiral and dwarf galaxies, 
their distribution is less clustered than that of optically selected galaxies.
HI surveys also significantly enhance the connectivity of large-scale 
filaments by filling in gaps populated by gas-rich galaxies.
\end{abstract}
\thispagestyle{plain}

\section{The HIPASS Galaxy Catalogs}
The HIPASS Bright Galaxy Catalog (BGC, Koribalski et al. 2004) contains 
the 1000 HI-brightest galaxies in the southern sky as obtained from the HI 
Parkes All-Sky Survey. These were selected by their HI peak flux density 
($S_{\rm peak} \ga 116$ mJy) measured in the spatially integrated HIPASS
spectrum. We note that a subset of $\sim$500 galaxies is complete in 
integrated HI flux density ($F_{\rm HI} \ga 25$ Jy\,km\,s$^{-1}$).
The HIPASS BGC is highly reliable (S/N $\ga$ 9) and contains accurate HI 
properties of all sources as well as optical identifications were available. 
Sources without cataloged optical (or infrared) counterparts, i.e. $\sim$10\% 
of the BGC, fall into two categories: (1) gas-rich galaxies in regions of high 
Galactic extinction or high stellar density, and (2) late-type dwarf and low 
surface brightness (LSB) galaxies. The latter are generally missed in optical 
catalogs because they are small in size or rather faint, but they are easily 
identified given the HI position. For details of the newly cataloged galaxies 
in the HIPASS BGC, 70\% of which lie in the Zone of Avoidance ($|b| < 
10^{\circ}$), see Ryan-Weber et al. (2001). We found only one extragalactic 
HI cloud, HIPASS J0731--69 ($M_{\rm HI} \sim 10^9$ M${\odot}$), located at 
a projected distance of 180 kpc from the asymmetric spiral galaxy NGC~2442 
(Ryder et al. 2001). 

\begin{figure}[!ht]
\plottwo{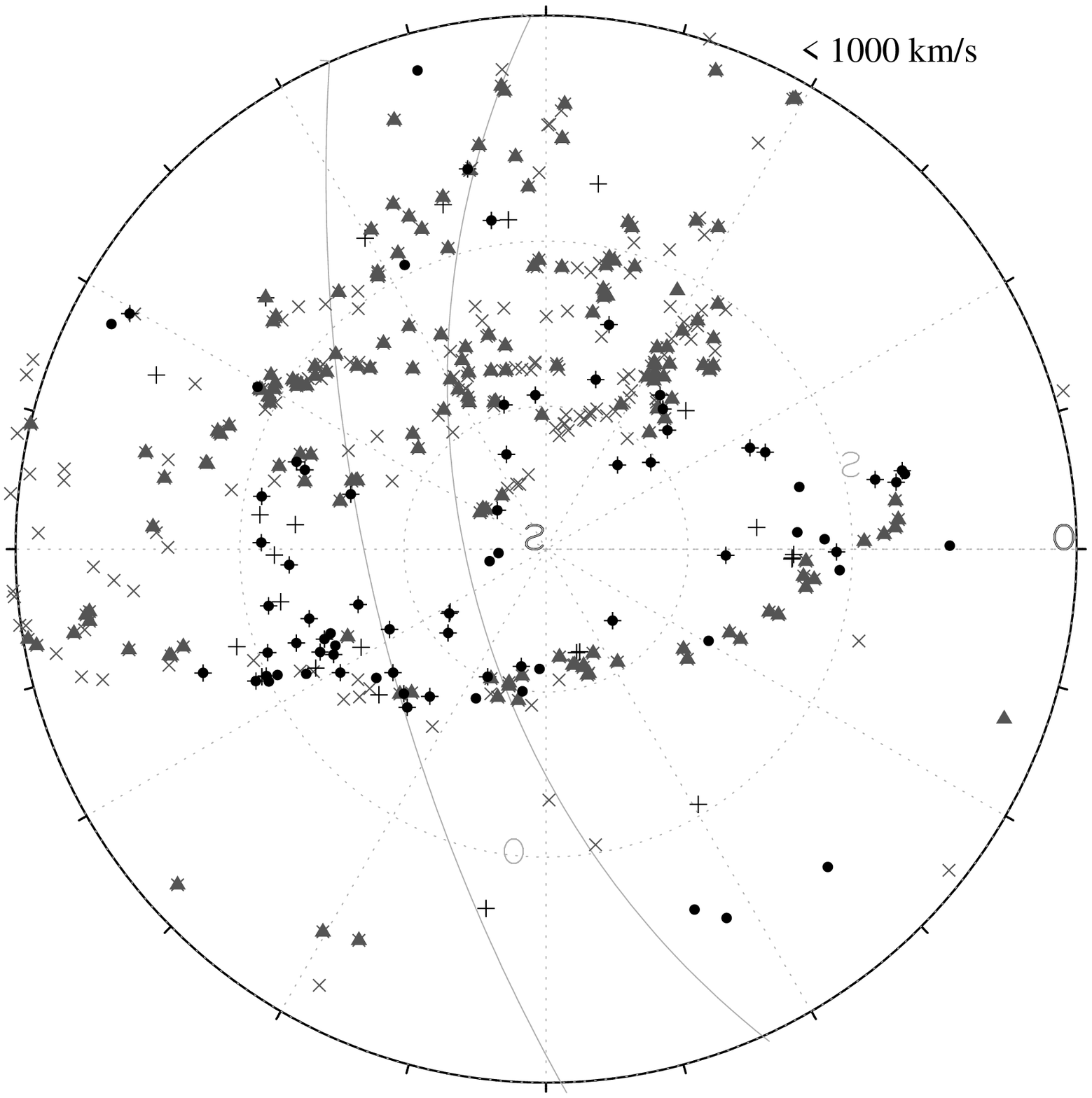}{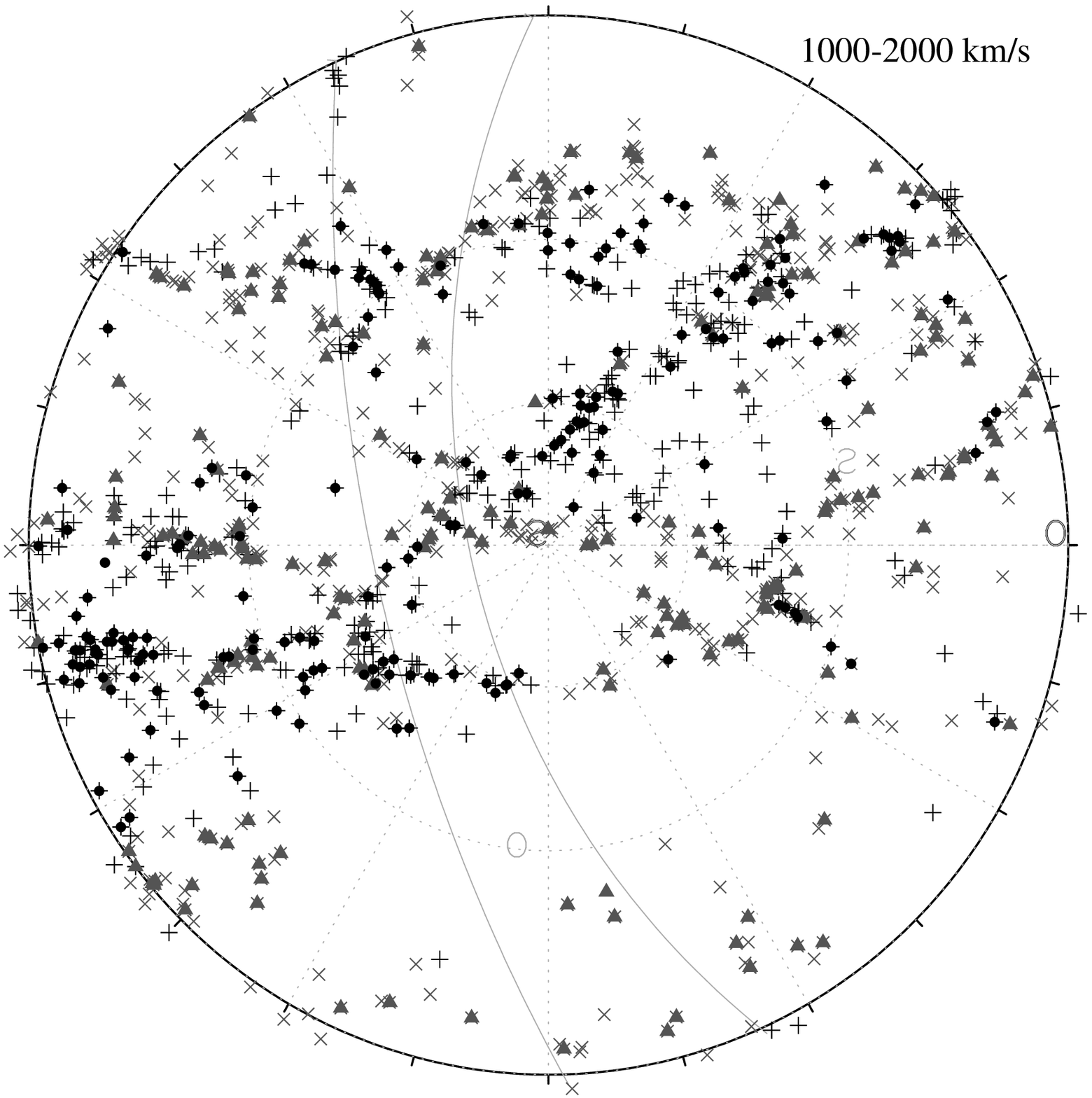}
\caption{Distribution of galaxies in the southern sky, divided into 
  several velocity ranges (indicated at the top right of each panel). 
  The filled black circles and triangles indicate HIPASS BGC sources, 
  while the plus signs and crosses indicate HICAT sources. Within each 
  velocity bin we distinguish galaxies in the lower half (circles and 
  plus signs) and the upper half (triangles and crosses). Galactic 
  latitudes of $b = \pm10^{\circ}$ are marked to emphasize the many new 
  structures seen across the (optical) Zone of Avoidance. We use a 
  stereographic projection onto the South Pole; Right Ascension tick 
  marks on the outer circle (declination $\delta(J2000) = 0^{\circ}$) 
  are given every hour, starting at the righthand side and increasing 
  counter-clockwise.}
\end{figure}
\setcounter{figure}{0}
\begin{figure}[!ht]
\plottwo{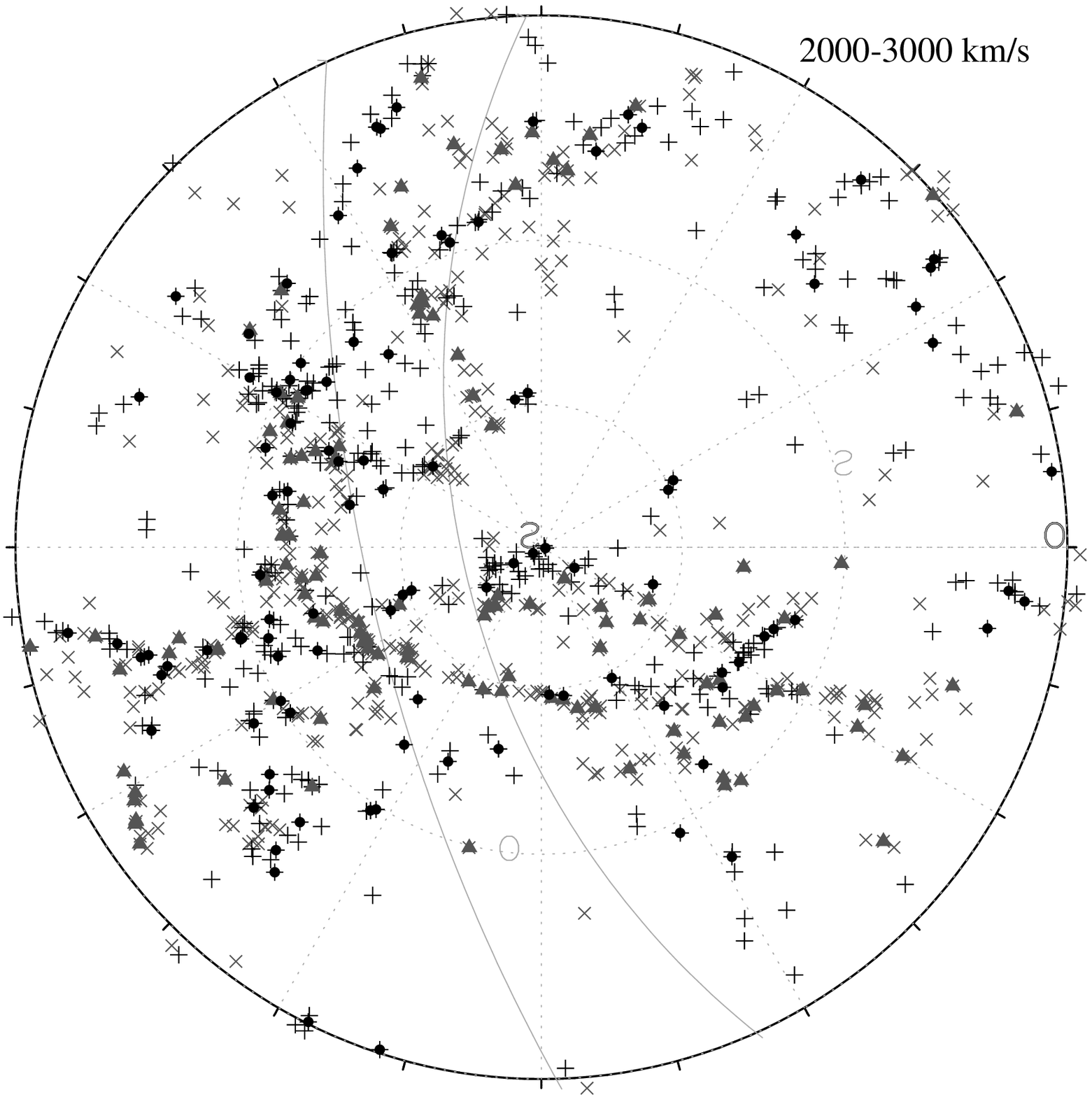}{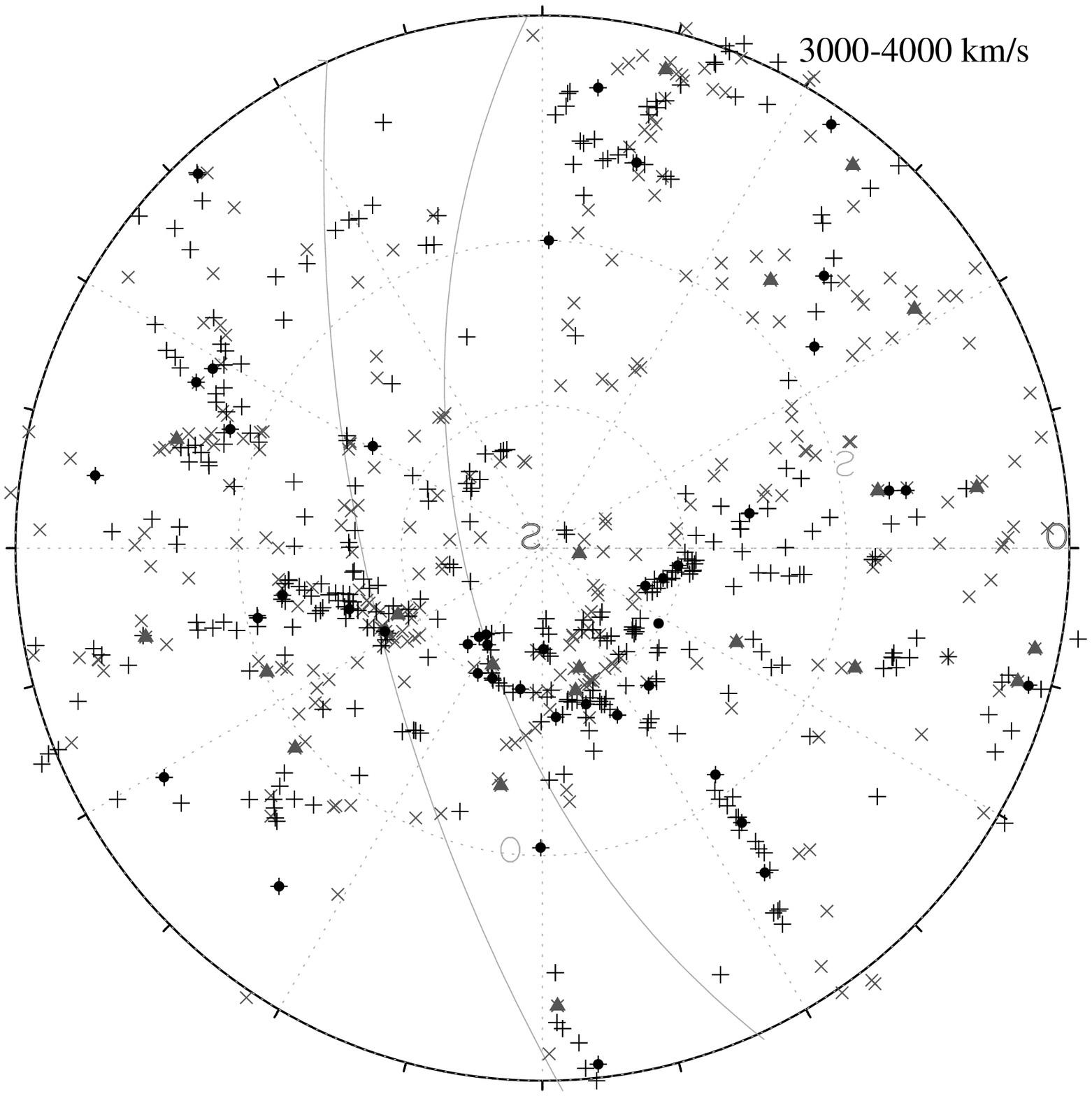}
\plottwo{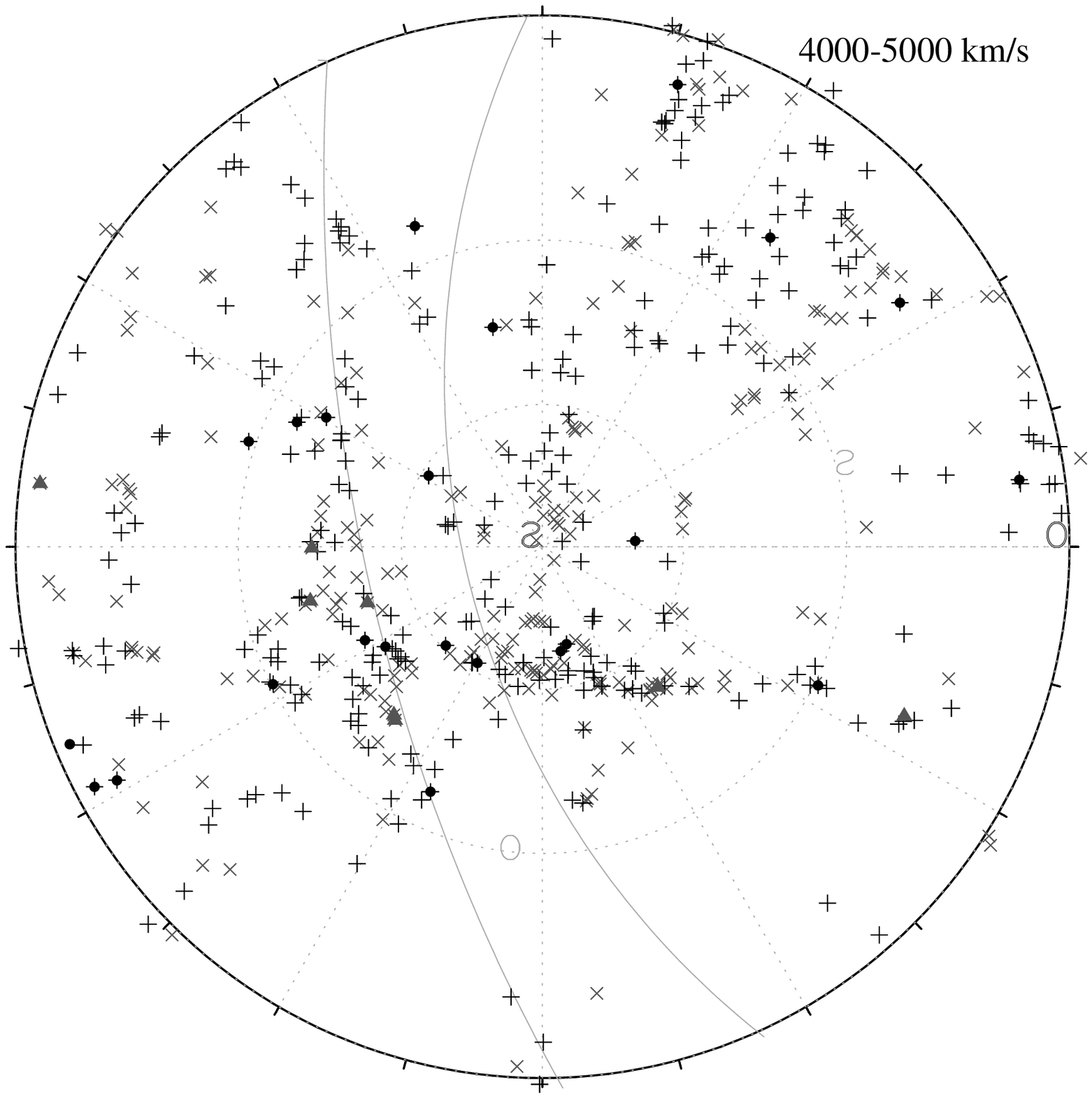}{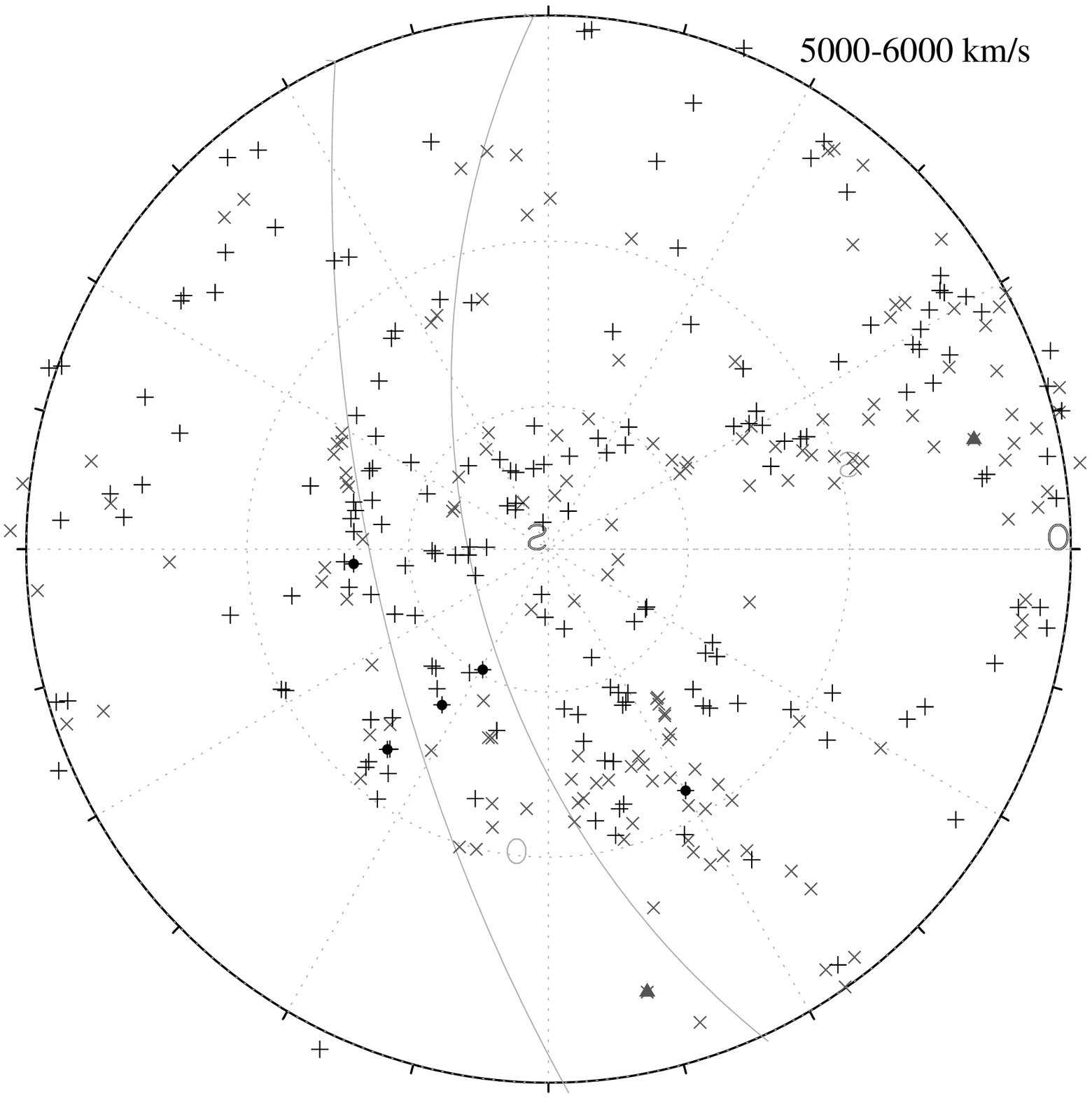}
\caption{--- continued.}
\end{figure}

Fig.~1 shows the distribution of all sources in the HIPASS BGC and HICAT,
divided into velocity bins of $\Delta v$ = 1000 km\,s$^{-1}$. Fig.~2 shows
the histogram of local group velocities in both catalogs. The deep HIPASS 
catalog (HICAT) contains 4315 HI sources, including a large fraction 
of the BGC. For the first time, we can trace local large-scale structures 
over the whole southern sky, unobscured by dust or foreground stars. A 
beautiful network of filaments, loops, groups and voids is revealed, creating 
a cellular web that gradually changes with redshift. In particular, HIPASS 
uncovers previously obscured galaxy structures across the Zone of Avoidance. 
The most striking feature in the first velocity bin ($<$1000 km\,s$^{-1}$) 
is the --- in this projection --- nearly horizontal, well defined narrow 
filament which corresponds to 
the near-side of the Supergalactic Plane (SGP). It is very long, reaching 
from the Sculptor Group on the righthand side all the way past the Cen\,A 
Group on the left, clearly continuing through the optical Zone of Avoidance
(ZOA). The voids on both sides of the plane help to delineate it so clearly. 
Also visible is the near-side of the Fornax Wall whose full extend is seen 
in the next velocity bin (1000 -- 2000 km\,s$^{-1}$), revealing another ZOA 
crossing. The criss-crossing of filaments (or walls) creates a remarkable 
cellular structure depicting regions of high galaxy density and voids. The 
local large-scale structures are further described in Section~3. Overall, we 
find that 
the filaments seen in HI appear well defined and continuous while optically 
outlined structures are generally much more clumpy (Koribalski et al. 2004). 
In other words, the correlation length of HI rich galaxies, most of which are 
late-type spiral galaxies (see Fig.~3) is much smaller than that of optically 
selected galaxies, which includes many more elliptical and early-type galaxies.
This decrease of clustering from early- to late-type galaxies (Giuricin et al. 
2001) affects our view of the local large-scale structure such that HI surveys
reveal many previously unknown structures and enhance connectivity within
known structures.

Fig.~4 also shows the distribution of galaxies on the southern sky, but here 
symbols indicate the galaxy morphological type. While optical identification
of the HIPASS BGC sources revealed $\sim$90\% cataloged galaxies (see 
Koribalski et al. 2004), cross-identification for HICAT is still under way.
The left side of Fig.~4 shows the optically identified BGC sources with their
mean morphological type as obtained from the Lyon Extragalactic Database 
(LEDA) as well as all newly cataloged 
sources. A preliminary match of all HICAT sources with velocities below 
4000 km\,s$^{-1}$ was also carried out using LEDA; the result is shown on
the right side of Fig.~4.

\section{HI Observations}
The HI Parkes All-Sky Survey (HIPASS) was carried out with the 21-cm multibeam 
receiver (Staveley-Smith et al. 1996) installed at the prime focus of the 
Parkes\footnote{The Parkes telescope is part of the Australia Telescope which 
 is funded by the Commonwealth of Australia for operation as a National 
 Facility managed by CSIRO.} 64-m telescope. HIPASS covers the whole sky up 
to declinations of $+25^{\circ}$. While the southern sky has 
been thoroughly searched  for HI sources, source finding in the northern data 
cubes is still under way. HIPASS covers a velocity range of $-1200 < cz < 
+12700$ km\,s$^{-1}$, at a channel spacing of 13.2 km\,s$^{-1}$. The velocity 
resolution is 18 km\,s$^{-1}$. After data reduction (see Barnes et al. 2001)
the gridded Parkes beam is typically 15\farcm5.
HIPASS data was obtained by scanning the telescope across the sky in 
8$^{\circ}$ strips in Declination ($\delta$). Five sets of independent scans 
were made of each region, resulting in a final sensitivity $\sim40$ 
mJy\,beam$^{-1}$ (3$\sigma$) which corresponds to a column density limit of 
about $6 \times 10^{18}$~cm$^{-2}$ (for objects filling the beam) and an HI 
mass limit of about $M_{\rm HI} = 10^6 \times D^2_{\rm Mpc}$ M$_{\odot}$ 
(assuming $\Delta v$ = 100 km\,s$^{-1}$). Note that the HI Parkes survey of 
the Zone of Avoidance (Henning et al.; these proceedings) is five times 
deeper.
\begin{figure}[!ht]
\plotone{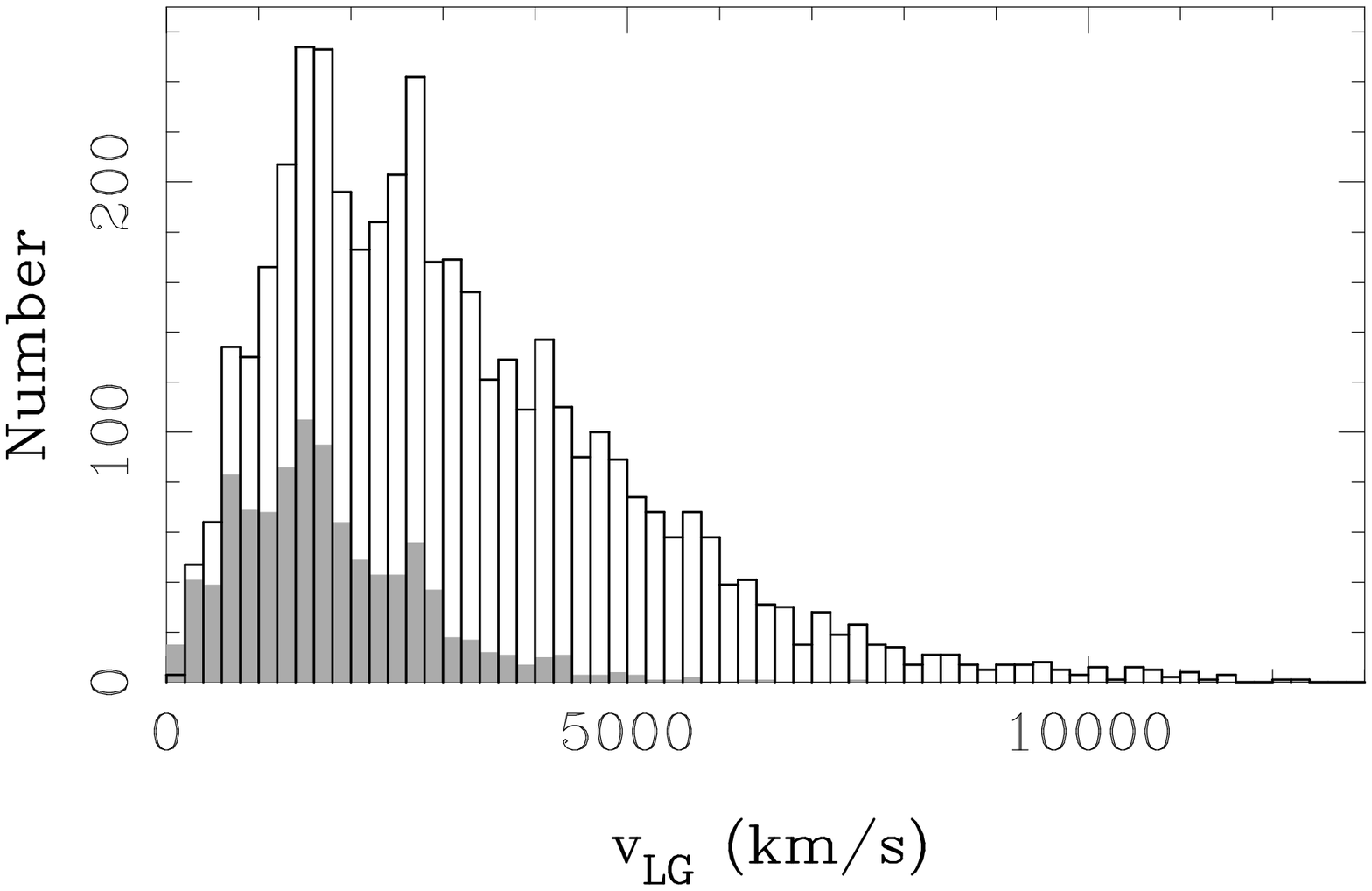}
\caption{Histograms of the Local Group velocities in the HIPASS BGC (grey)
  and HICAT (white).}
\end{figure}

\begin{figure}[!ht]
\plottwo{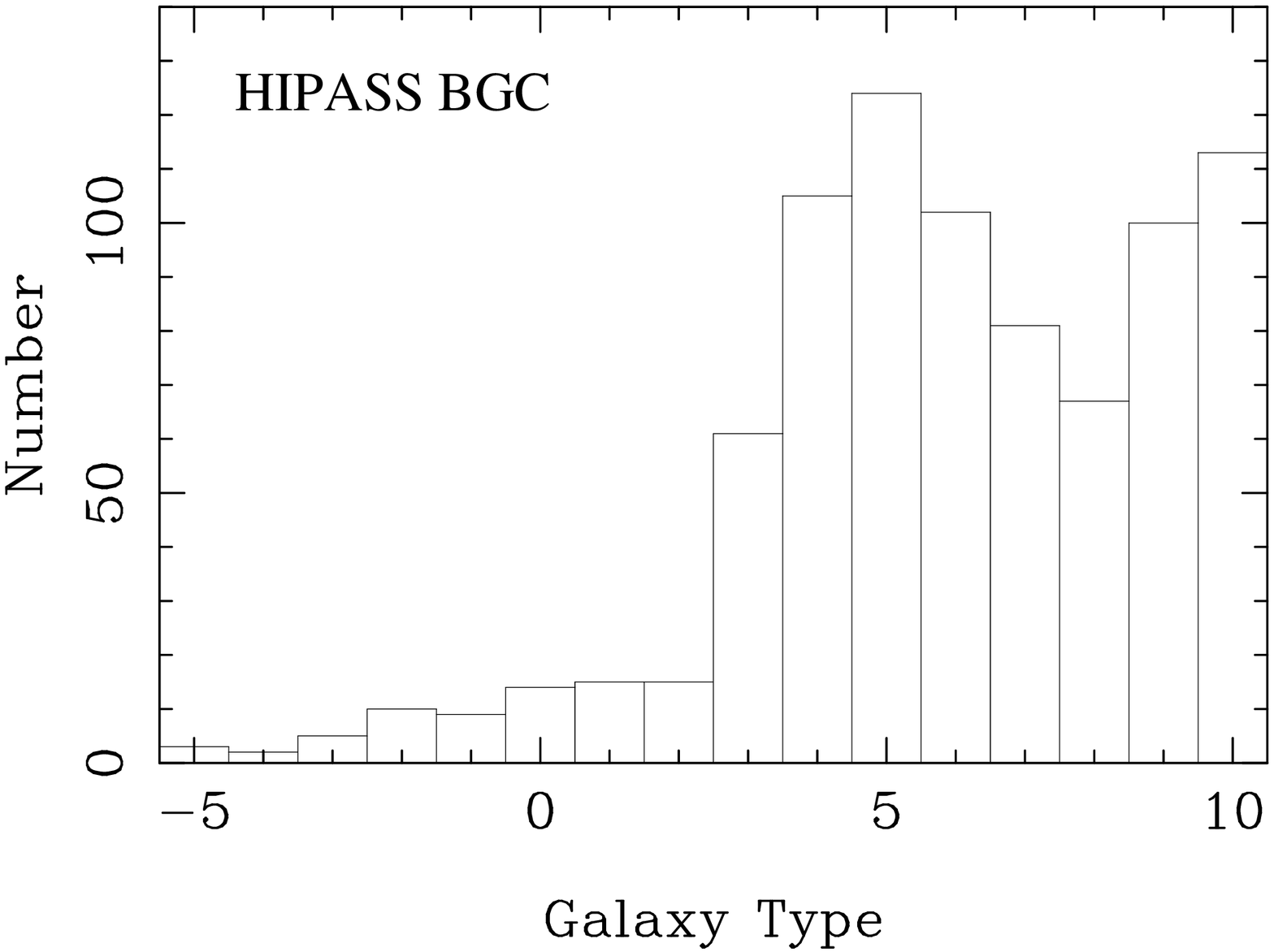}{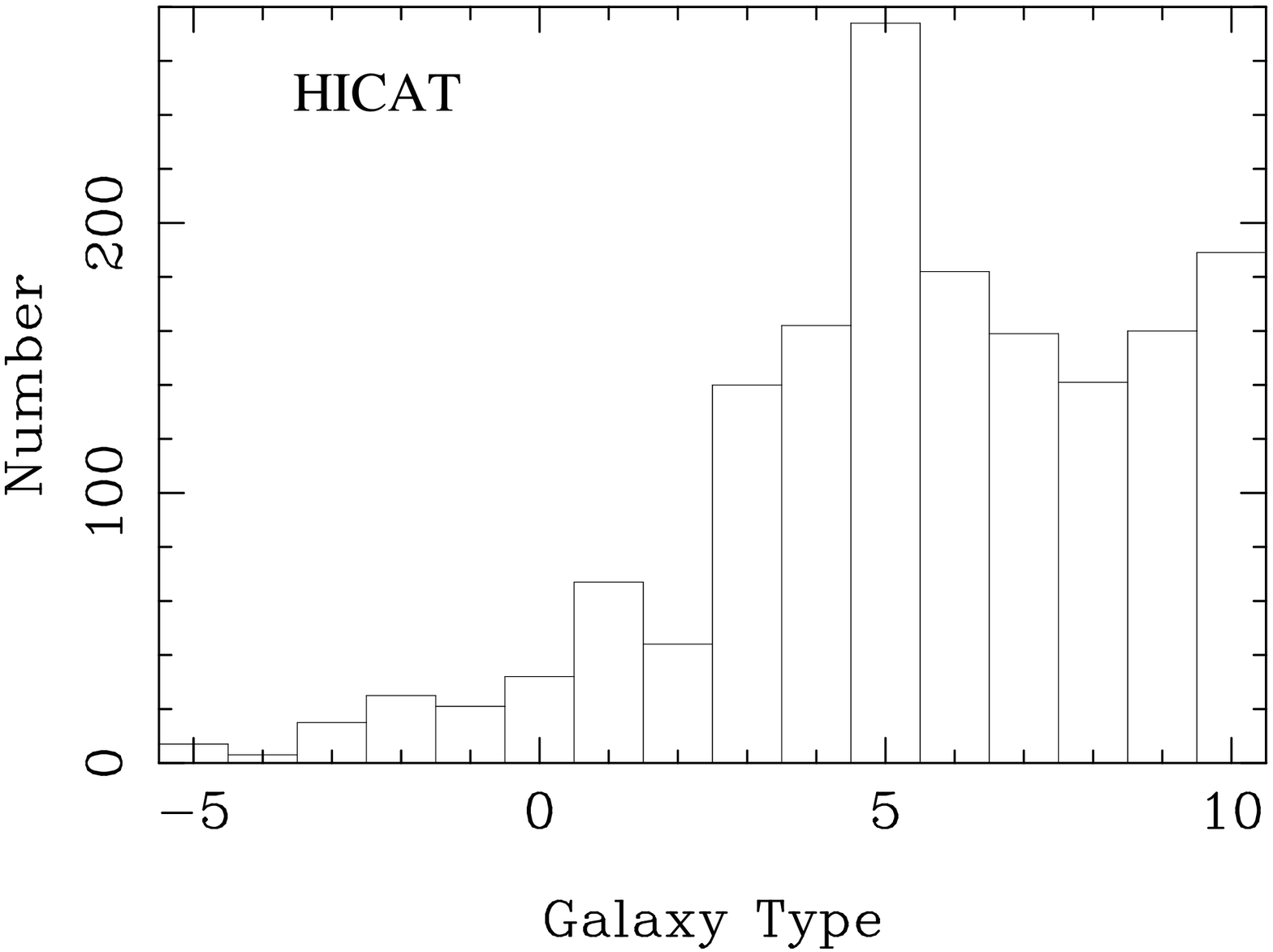}
\caption{Histograms of galaxy morphological types in the HIPASS BGC (left)
  and HICAT (right), for all optically identified galaxies as obtained from 
  LEDA. Optical identification of HICAT sources is not yet complete and we
  only show a subset of galaxies with velocities less than 4000 km\,s$^{-1}$.
  The numerical galaxy types (de Vaucouleurs et al. 1991) are: --5 (E), --2 
  (S0), 1 (Sa), 3 (Sb), 6 (Sc), 8 (Sd), 9 (Sm), and 10 (Irr), plus intermediate
  types; note that 5 (S?) corresponds to galaxies with uncertain spiral type. 
  In both samples there is, as expected, an abrupt decline in galaxy numbers 
  for morphological types earlier than Sb.}
\end{figure}

The advantages of HI surveys are numerous: with 21-cm multibeam receivers 
such as the Parkes 13-beam system we can now cover large volumes out to 
several hundreds of Mpc. We simultaneously measure HI source position and 
velocity, resulting in accurate systemic velocities, velocity widths and HI 
masses, as well as --- supplemented by, e.g., the optical galaxy inclination 
--- estimates of the rotation speed and therefore the total dynamical mass. 
HI observations are unaffected by dust extinction or foreground stars, 
resulting in an unobscured view of the nearby large-scale structure. Since 
LSB and dwarf irregular galaxies tend to be gas rich, these are easily 
detected in HI surveys. We emphasize that HI and optical (infrared) surveys 
are complementary, as the former favours gas-rich spiral and irregular 
galaxies while the latter are biased toward bright spiral and elliptical 
galaxies. The disadvantages of current HI surveys are that existing telescopes 
limit us to study the nearby galaxy structures and that single dish surveys
have relatively low angular resolution, which means increasing source confusion
at higher redshifts.

\begin{figure}[!ht]
\plottwo{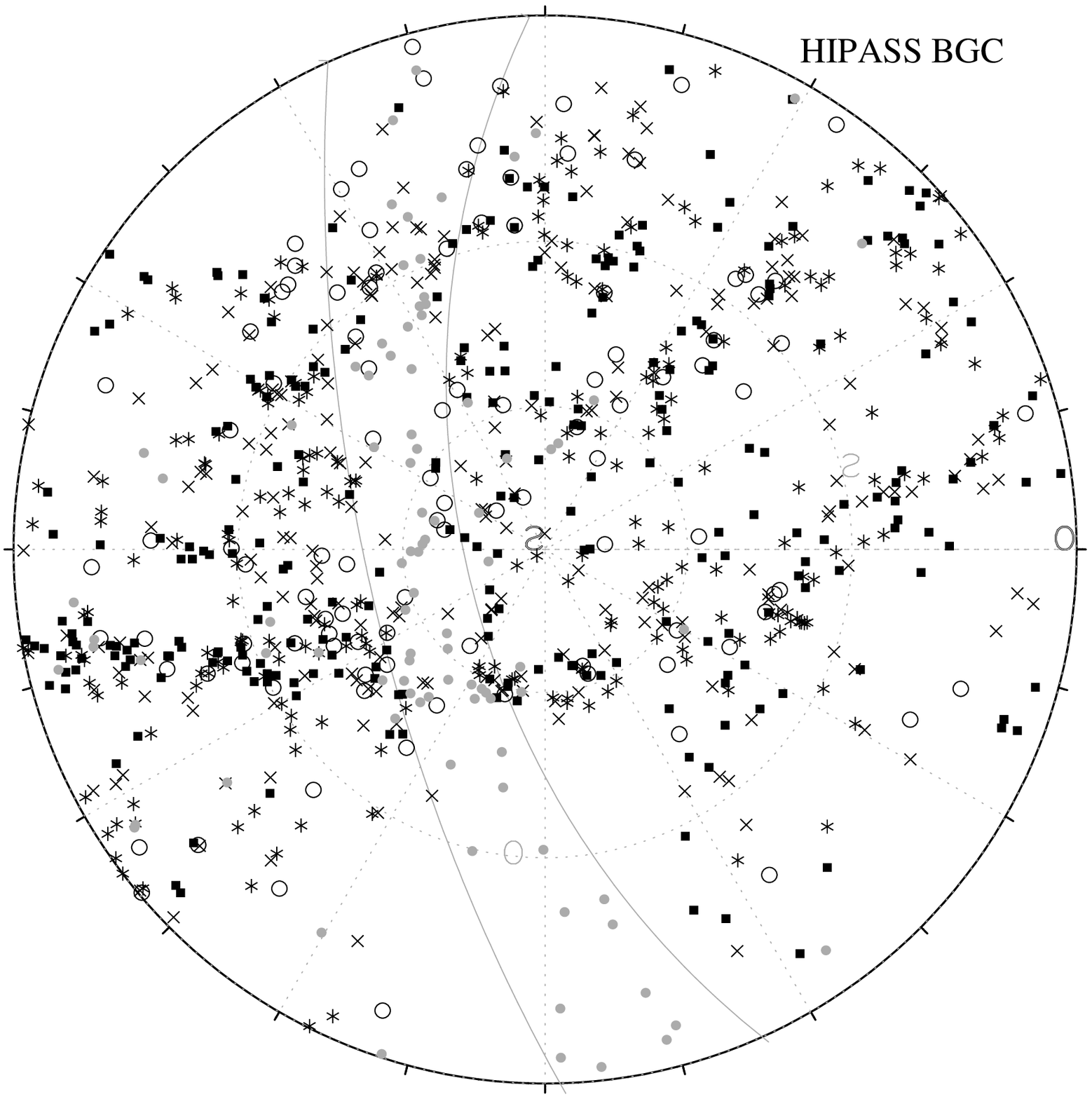}{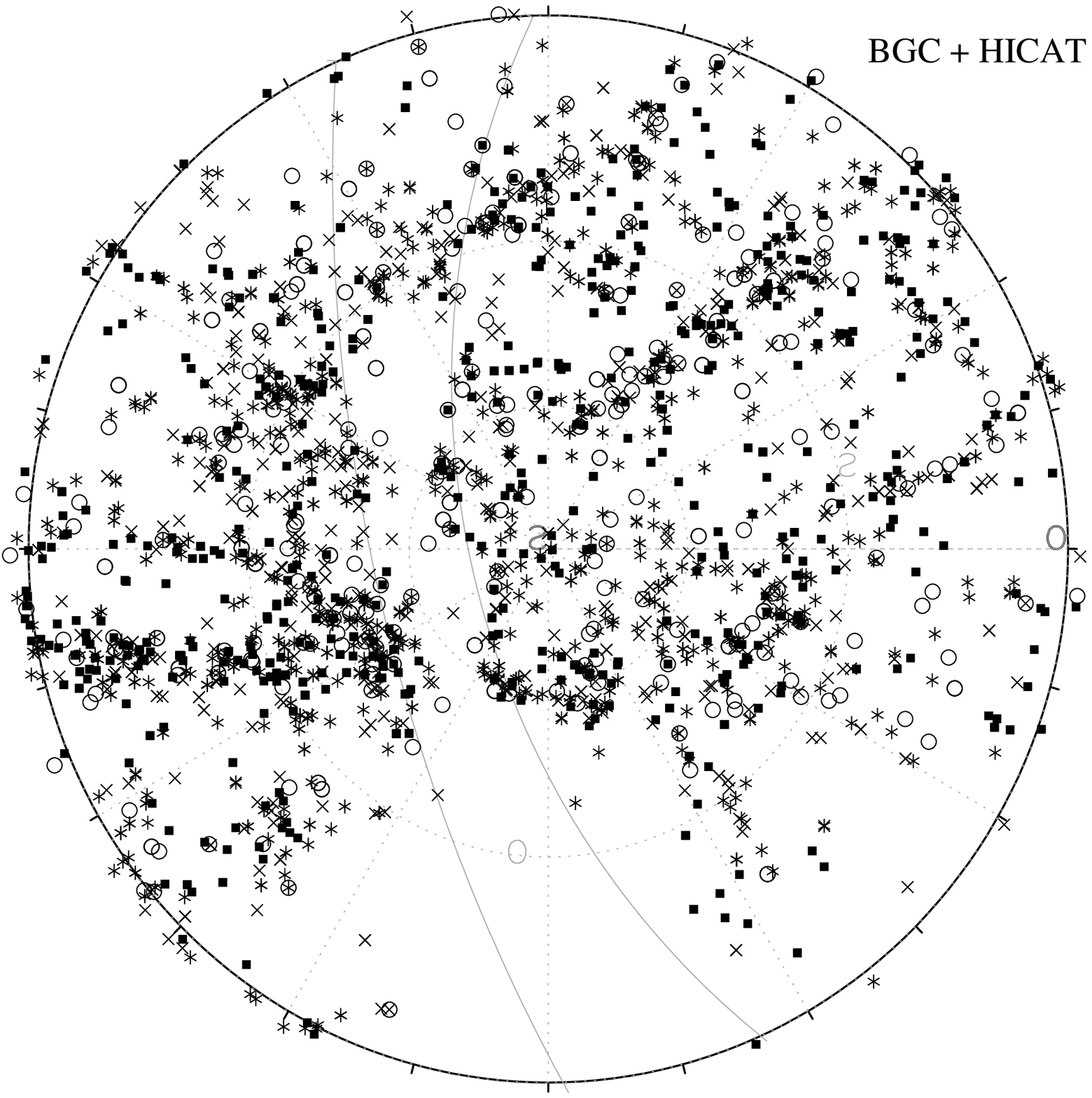}
\caption{Distribution of galaxies in the southern sky. The symbols indicate 
  mean morphological types (when available) as obtained from LEDA: type $<$ 
  2.5 (circles), 2.5--5.0, (crosses), 5.0--7.5 (stars), and 7.5--10 (black 
  squares), roughly corresponding to Sab \& earlier, Sb, Sc, and Sd/m \& Irr, 
  respectively.
  {\bf (Left)} Optically identified galaxies in the HIPASS BGC plus 
       newly cataloged galaxies (filled grey circles). 
  {\bf (Right)} Optically identified galaxies in the HIPASS BGC and HICAT. 
  The latter are only included if reliably identified in position and 
  velocity with $v_{\rm sys} < 4000$ km\,s$^{-1}$.}
\end{figure}

\section{Local Large-Scale Structure}
The galaxies detected in HIPASS give us the first view of the local Universe 
uninhibited by the foreground stars and dust from our own Galaxy. Numerous 
known large-scale structures can now be fully traced across the southern sky, 
the most prominent being the Supergalactic Plane, followed by the Fornax and 
Hydra Walls. Their extensions and connections across the optical Zone of 
Avoidance are clearly revealed in HIPASS (see Fig.~1). In addition, HIPASS
provides a less clustered view of the local Universe by predominantly detecting
late-type spiral and dwarf galaxies. These appear to trace large-scale 
structures in a much more homogeneous way than optically selected galaxies.
The SGP, Fornax and Hydra Walls create a giant three-pronged structure
resembling the footprint of a dinosaur, as pointed out by Lynden-Bell (1994).
HIPASS also allows a much improved delineation of voids as, for example, in 
the case of the Local Void.

At velocities below 1000 km\,s$^{-1}$ the most prominent features are the 
Supergalactic Plane (SGP), the Local Void next to the Puppis filament, 
the near side of the Fornax Wall and the Volans Void.
The Supergalactic Plane (SGP) contains the greatest concentration of nearby 
groups and clusters of galaxies in the local Universe (see Lahav et al. 2000, 
and references therein). The SGP is very evident in HIPASS which reveals a 
distinct and continuous filamentary band --- without a gap due to the ZOA --- 
across the whole southern sky ending on the righthand side in the Southern 
Virgo Extension. The latter becomes quite prominent in the second velocity
bin. The width of the SGP is no more than $\sim10^{\circ}$; deviations of 
up to $10^{\circ} - 15^{\circ}$ from the supergalactic equator are apparent.  
HIPASS is also more stringent in defining the void sizes. The dominant void 
in the nearby Universe is the Local Void which can be seen as the empty 
region below the SGP. 

At velocities between 1000 and 2000 km\,s$^{-1}$ we find the known clumpings 
of Fornax, Eridanus and the Southern Virgo Extension as well as two new galaxy 
groups which are not at all visible in the optical: the NGC~4038 and NGC~5084 
Groups. For a detailed description see Koribalski et al. (2004).

In the next velocity bin (2000 -- 3000 km\,s$^{-1}$) one prominent filament 
which stretches over a major fraction of the southern sky stands out clearly. 
It can be traced from the Indus and Pavo clusters crossing the ZOA to the 
Centaurus cluster in a linear structure, called the Centaurus Wall by Fairall 
(1998), at a slight angle with respect to the SGP. From there it bends over 
to the Hydra and Antlia clusters and folds back across the ZOA through Puppis 
where yet another spiral-rich new galaxy group is uncovered, Puppis~G2.
Also outstanding is the large-scale underdensity known as the Eridanus Void.

\acknowledgments
This work would not have been possible without the dedication of many 
HIPASS and ZOA team members. I particularly like to acknowledge Renee
Kraan-Korteweg for her help in identifying known and new structures,
filaments and groupings in the southern sky. Also, thank you very much 
to our hosts, Tony Fairall and Patrick Woudt, for organizing such an 
interesting conference at a most remarkable and beautiful location.



\begin{thebibliography}{}
\bibitem[2001]{barnes:etal}
Barnes, D.G., Staveley-Smith, L., et al. 2001, MNRAS 322, 486 
\bibitem[1998]{fairall}
Fairall, A.P. 1998, Large-Scale Structures in the Universe, Chichester, 
  Praxis Publishing Ltd.
\bibitem[2001]{giuricin:etal}
Giuricin, G., Samurovic, S., Girardi, M., Mezzetti, M., Marinoni, C. 2001,
  ApJ 554, 857
\bibitem[2004]{koribalski:etal}
Koribalski, B.S., Staveley-Smith, L., Kilborn, V.A. et al. 2004, AJ 128, 16
\bibitem[2000]{lahav:etal}
Lahav, O., et al. 2000, MNRAS 312, 166
\bibitem[1994]{lynden-bell}
Lynden-Bell, D. 1994, in ``Mapping the Hidden Universe'', the ASP Conf. Series
  Vol. 67, eds. C. Balkowski \& R.C. Kraan-Korteweg, p.\,289
\bibitem[2004]{meyer:etal}
Meyer, M. et al. 2004, MNRAS 350, 1195
\bibitem[2002]{ryan-weber:etal}
Ryan-Weber, E., Koribalski, B.S., et al. 2002, AJ 124, 1954
\bibitem[2001]{ryder:etal}
Ryder, S., Koribalski, B.S., et al. 2001, ApJ 555, 232
\bibitem[2001]{ryder:etal}
Staveley-Smith, L., et al. 1996, PASA 13, 243
\bibitem[1991]{vaucouleurs:etal}
de Vaucouleurs, G., et al.  1991, Third Reference Catalogue of Bright
   Galaxies, New York, Springer Verlag
\end{thebibliography}
\end{document}